\documentclass[
  ,final            
  ]
  {aipproc}

\layoutstyle{6x9}
\usepackage{bm}

\newcommand{\bkt}{\bm k_T}

\newcommand{\ba}{\begin{eqnarray}}
\newcommand{\ea}{\end{eqnarray}}
\newcommand{\nn}{\nonumber}
\newcommand{\beq}{\begin{equation}}
\newcommand{\eeq}{\end{equation}}
\newcommand{\bc}{\begin{center}}
\newcommand{\ec}{\end{center}}
\newcommand{\bmp}{\begin{minipage}}
\newcommand{\emp}{\end{minipage}}

\newcommand{\slsh}[1]{\mbox{$\not\! #1$}}

\newcommand{\psibar}{\overline{\psi}}
\newcommand{\la}{\langle}
\newcommand{\ra}{\rangle}
\newcommand{\amp}[1]{\la #1 \ra}

\newcommand{\text}{\hbox}

\newcommand{\Tr}{\text{Tr}}

\begin{document}

\title{Hadron structure from $\gamma^* \, p$ scattering: interpreting hadronic
matrix elements
\footnote{Presented at the 19th European Conference on Few-Body Problems 
in Physics, Groningen, The Netherlands, August 23-27, 2004}\\[-3 mm]}

\author{Dani\"{e}l Boer\\[-3 mm]}{address={Department of Physics and 
Astronomy, Vrije Universiteit Amsterdam\\
De Boelelaan 1081, NL-1081 HV Amsterdam,
The Netherlands\\[-3 mm]}
}

\begin{abstract}
Hadron structure from high-$Q^2$ $\gamma^*\, p$ scattering processes
is often 
expressed in terms of hadronic matrix elements of nonlocal operators. 
Properly defining and interpreting these quantities is very important 
in light of experiments aiming to
extract transverse momentum dependent parton distributions or   
generalized parton distributions. The current status will be 
reviewed, including recent developments concerning Wigner distributions.\\[-10 mm]
\end{abstract}

\maketitle


\section{Introduction}

The hard $\gamma^* \, p$ scattering processes to be considered are inclusive 
deep inelastic scattering (DIS), semi-inclusive DIS and deeply virtual 
Compton scattering (DVCS). In these processes one probes parton densities, 
transverse momentum dependent parton distributions (TMPDs) and generalized 
parton distributions (GPDs), respectively. Properties 
of these quantities will be reviewed, with emphasis on 
interpretation, since having a clear interpretation is
crucial for motivating experiments. The various distributions all carry 
different information about hadron structure, but 
can be viewed as different reductions of one underlying quantity, a quantum
phase space (Wigner) distribution.\\[-10 mm] 

\section{Hard inclusive $\gamma^*$ {\it \lowercase{p}} processes} 
Inclusive DIS ($e\, p \to e' \, X$) is a process of two scales, provided by 
the hadron momentum $P$ and the virtual photon momentum
$q$: $P^2 = M^2$ and $Q^2=-q^2$, such that $Q^2 \gg M^2 $. 
The hard scale $Q$ serves to separate hard (perturbative) from soft
(nonperturbative) parts of the process, i.e.\ one can apply factorization: 
\begin{center}
$\sigma (\gamma^* \, p \to X) \propto$
\begin{minipage}{4.5 cm}
\includegraphics[width = 4.5 cm]{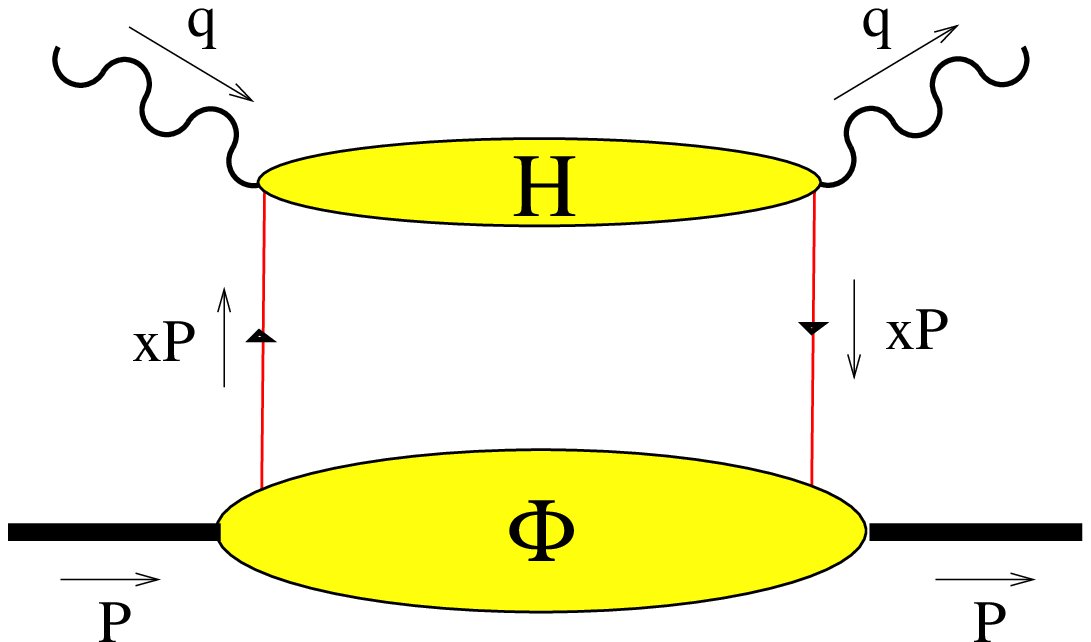}
\end{minipage} $=   
\int d x \, \text{Tr} \left[ H_{\mu\nu}(x) {\Phi (x)} 
\right] + {\cal O}\left(1/Q\right)
$
\end{center}
In DIS one only encounters functions of lightcone momentum fractions
($x= k^+/P^+$ is the fraction of light-cone momentum of a quark $(k^+)$
inside a hadron $(P^+)$). Here  
\beq
{\Phi_{ij}(x)} =  
\int \frac{d \lambda}{2\pi}\, e^{i \lambda x}\, \langle P \vert \; 
{\overline \psi_j(0) \, {\cal L}[0,\lambda] \, \psi_i(\lambda n_-)} \;
\vert P \rangle, \qquad \quad (n_-^2=0)
\eeq
is an operator matrix element (OME) 
of a nonlocal lightcone operator and 
\beq
{{\cal L}[0,\lambda]} = {\cal P} \exp \left(-ig\int_0^{\lambda} 
d\eta \, A^+ (\eta n_-)\right) \stackrel{{A^+ =0}}{ 
\longrightarrow} 1 
\eeq
This {path-ordered exponential (link)} is not inserted by hand, but 
{\em derived} \cite{Efremov:1978xm}. 
For an unpolarized hadron $\Phi(x)$ can be parametrized
by one function (at leading twist):
\beq
{\Phi(x)} = \frac{1}{2} {f_1(x)} \mbox{$\not\! P\,$}.
\label{Phiparamx}
\eeq
This function $f_1^{{q}} (x) \equiv {q}(x)$ can be written in terms of the
good field $\psi_+ = \frac{1}{2} \gamma^- \gamma^+ \psi$ as
\beq
q(x) = \frac{1}{\sqrt{2}} \; \sum_n \; \delta(P^+ +xP^+ -p_n^+) \; 
\left|{\amp{n | \; {{\cal
L}[\infty,0] \; \psi_+ (0)} \; 
|P }}\right|^2, 
\eeq
which lends $q(x)$ the interpretation of probability 
of finding a quark of flavor $q$ in the proton, with a light-cone momentum
fraction $x$. Taking Mellin moments $\int dx \; {x^N} \; q(x)$, leads to
local OMEs (which can be evaluated on the lattice). For example, $N=0$ yields
the number of quarks minus antiquarks: $\int_{-1}^1 dx \; q(x) = 
\amp{P | \; {\psi_+^\dagger(0) \psi_+(0)} \; | P }/(\sqrt{2} P^+)$. 
 
Semi-inclusive DIS (SIDIS), ${e \, p \to e' \, {\pi} \, X}$, is a three-scale
process, where next to $M$ and $Q$ also the transverse momentum
of the measured final-state hadron (a pion for definiteness) sets a scale: 
$|\bm P^{\pi}_{\perp}|$ with $|\bm P^{\pi}_{\perp}|^2 \ll Q^2$. 
This results in a different factorization theorem (recently discussed in Ref.\
\cite{Ji:2004wu}, based on methods developed in Ref.\ \cite{Collins:1981uk}). 
One needs to 
include parton transverse momentum {$\Phi(x) \rightarrow \Phi(x,\bkt)$}
(discussed in Refs.\ \cite{Soper:1976jc,Ralston:1979ys} and by many others), 
which leads to 
transverse momentum dependent parton distribution functions (TMPDs).
TMPDs for {unpolarized} hadrons are defined as \cite{Boer:1998nt}
\beq
\Phi(x,\bm{k}_T) = 
{f_1(x ,\bm k_T^2)}\, \frac{\slsh{P}}{2} + 
{h_1^\perp (x,\bm{k}_T^2)} \, 
\frac{i \slsh{k_T} \slsh{P}}{2M}.
\eeq
Upon integration over transverse momentum one retrieves Eq.\
(\ref{Phiparamx}). {$\Phi(x,\bkt)$} is a matrix element of operators
that are nonlocal {\em off the lightcone\/} 
\beq
\Phi(x,\bm{k}_T) =  \text{F.T.} \,
\amp{P| \; \psibar(0) \, {{\cal L}[0,\xi]} \, 
 \psi(\xi) \; |P}\bigg|_{{\xi = (\xi^-,0^+,\bm{\xi}_T)}}.
\eeq
Like for $\Phi(x)$, the link ${{\cal L}[0,\xi]}$ can be {derived} as discussed
in Refs.\ \cite{Boer:1999si,Belitsky:2002sm,Boer:2003cm}.  
In many respects $\Phi(x,\bkt)$ 
is similar to $\Phi(x)$ (it defines momentum distributions), 
but the link structure leads to considerable differences, of which two will be
mentioned here. 

A nonzero function $h_1^\perp $ means that 
the transverse polarization $S_T^q$ of a {noncollinear} quark inside an
unpolarized hadron in principle can have a preferred direction.
This implies an {intrinsic handedness}, e.g.\ in
the infinite momentum frame (IMF): 
${S_{T}^q \sim P_{\text{hadron}} \times k_{\text{quark}}}$.
At first sight such handedness 
appears to {violate time reversal invariance}
(following an argument of Collins \cite{Collins:1993kk}), but a 
model calculation by Brodsky, Hwang and Schmidt \cite{Brodsky:2002cx} 
implied otherwise. This is precisely due to the link structure. 
The proper gauge invariant definition of TMPDs
{in SIDIS contains a future
pointing Wilson line}, whereas in Drell-Yan (DY) it is past pointing.  
As a consequence there is a {\em calculable\/} process dependence
\cite{Collins:2002kn}: 
${(f_{1})_{{\rm DIS}}} = {(f_{1})_{{\rm DY}}}$, but  
${(h_{1}^\perp)_{{\rm DIS}}} =  {- (h_{1}^\perp)_{{\rm DY}}}$. 
More complicated processes \cite{Bomhof:2004aw} still require further
study.  

Another consequence of the link structure is that sometimes one is
dealing with intrinsically nonlocal lightcone OMEs, such as
\beq
h_{1}^{\perp {(1)}}(x) \equiv \int d^2 \bm{k}_T
{\frac{\bm{k}_T^2}{2M^2}} h_{1}^{\perp } (x, \bm{k}_T^2) 
\stackrel{A^+ =0}{\propto} \text{F.T.} \,
\amp{P | \; \psibar (0) \; 
{\int_{-\infty}^{\infty} \! d\eta^- \; F^{+\alpha} (\eta^-)}\; \Gamma \; 
\psi(\xi^-) \; | P},
\eeq
for which 
Mellin moments do {\em not\/} yield local OMEs (hampering a lattice 
evaluation).\\[-8 mm]

\section{Hard exclusive $\gamma^*$ {\it \lowercase{p}} processes}

To get a handle on {orbital angular momentum} of quarks, 
Ji proposed \cite{Ji:1996ek} 
to use {Deeply Virtual Compton Scattering}: $\gamma^* \, p \to \gamma \,
p'$.  
This involves Generalized Parton Distributions, which are 
off-forward, nonlocal lightcone OMEs
\ba
\lefteqn{\int \frac{d \lambda}{2\pi}\, e^{i \lambda x}\, \langle {P'}
\vert \;  
{\overline \psi(-\lambda/2) \, \gamma^+ {\cal L}[-\lambda/2,\lambda/2] 
\, \psi(\lambda/2)} \; \vert {P} \rangle =} \nn\\
 && \mbox{} \hspace{3 mm} {H_q(x,\xi,t)} \, 
\bar{u}(P') \gamma^+
u(P) + {E_q(x,\xi,t)} \, 
\bar{u}(P') \frac{i \sigma^{+ \nu} \Delta_\nu}{2M_N} u(P), 
\ea
with {$\Delta= P'-P$}, {$\xi = - \Delta^+/({P'}^+ + P^+)$} 
and {$t=\Delta^2$}. 
{GPDs encompass both parton distributions and form factors.
In the forward limit: $H_q(x,0,0) = q(x)$ and $E_q(x,0,0) \equiv E_q(x)$ (the
latter function is not accessible in DIS), which determine the total and
orbital {angular momentum} of quarks in a proton, $J_q(x)$ and $L_q(x)$ 
\cite{Hoodbhoy:1998yb},
\beq
{J_q(x)} = {\frac{1}{2} x \left[ q(x) + E_q(x) \right]},\quad 
{L_q(x)} = {J_q(x)}- \frac{1}{2}{\Delta q(x)},
\eeq
together with {$\Delta q(x)$}, the quark 
helicity distribution ($\Tr \left[\Phi(x) 
{\gamma^+ \gamma_5} \right] \sim {\lambda} {\Delta q(x)}$). 

The reduction to form factors goes via $x$-integration:
\beq
\int dx \; H(x,\xi, \Delta^2) = {F_1} (\Delta^2), \quad
\int dx \; E(x,\xi, \Delta^2) = {F_2} (\Delta^2),
\eeq
where ${F_1}$ and ${F_2}$ are the usual {Dirac and Pauli form factors}.
Like form factors, GPDs are best interpreted by taking Fourier transforms
(F.T.).  
In that way GPDs yield a more complete picture of
momentum {\em and} {spatial} {distributions of partons}
\cite{Burkardt:2000za,Ralston:2001xs,Belitsky:2002ep,Diehl:2002he}, albeit 
a frame dependent picture.
Two standard choices are the IMF and the Breit frame. 

In the IMF ($P_z \to \infty$) one effectively has 
{localization in the transverse directions}, which leads to a  
{2-D position space interpretation}
(information along the z-axis is integrated over). It allows to define 
the charge distribution in impact parameter space 
\beq
\rho(\bm{b}_\perp) \equiv \frac{1}{2P^+} 
\langle {P^+, \bm{R}_\perp = 0} \vert {J^+}(0^-,0^+,
{\bm{b}_\perp})
\vert {P^+, \bm{R}_\perp = 0} \rangle,
\eeq
where $\vert {P^+, \bm{R}_\perp = 0} \rangle$ is the proton state localized 
in the $\perp$ direction. One can show that 
\beq
\rho(\bm{b}_\perp) = \int \frac{d^2 \bm{\Delta}_\perp}{(2\pi)^2} \; 
e^{i \bm{\Delta}_\perp \cdot \bm{b}_\perp} \; {F_1(-\bm{\Delta}_\perp^2)}. 
\eeq
Hence, the {Dirac form factor} is the F.T. of the charge distribution in the
transverse plane.  

Taking this IMF point of view for GPDs leads to ($\int \! dz \to \xi=0$):
\beq
{H_q(x,0,- \bm{\Delta}_\perp^2) = \text{2-D} \ 
\text{F.T.} \ q(x, \bm{b}_\perp)}, 
\eeq
$\bm{b}_\perp$ is measured w.r.t.\ $\bm{R}_\perp^{CM} \equiv \sum_i x_i
\bm{r}_{\perp i}$: the `{transverse center of longitudinal momentum}'.
This {$q(x, \bm{b}_\perp)$ has an interpretation as a {\em density}},
just like $q(x)$ \cite{Burkardt:2000za,Soper:1976jc} 
\beq
q(x, \bm{b}_\perp) = \int \frac{d \lambda}{2\pi}\, e^{i \lambda x}\, 
\langle {P^+, \bm{R}_\perp = 0} \vert\;  
\overline \psi(-\lambda/2,{\bm{b}_\perp}) \, \gamma^+ \, {\cal L}
\, \psi(\lambda/2,{\bm{b}_\perp}) \; 
\vert {P^+, \bm{R}_\perp = 0} \rangle.
\eeq
\beq
\text{Note:} \quad 
{\int dx} \; q(x,\bm{b}_\perp) = \rho(\bm{b}_\perp), \quad  
{\int d^2\bm{b}_\perp} \;  q(x, \bm{b}_\perp)= q(x). 
\eeq

Alternatively, in the Breit frame 
($\vec{P}^{\; \prime} = - \vec{P}$ and $t=-\vec{\Delta}^{\;
2}$) the {3-D} proton 
charge distribution is the F.T. of the Sachs electric form factor $G_E$ 
\cite{Sachs:1960,Sachs:1962}
\beq
\text{3-D} \ \text{F.T.} \, {\rho(\vec{r})} 
= {\amp{\vec{\Delta}/2 | J^0(0) | -\vec{\Delta}/2}/(2M_N)} 
\propto {G_E(t) = F_1(t) + \frac{t}{4 M_N^2} F_2(t)}. 
\eeq
This Breit frame point of view for GPDs leads to ($\xi = \Delta^z/
(2P^0)$, $t = -\vec{\Delta}^{2}$) \cite{Ji:2003ak,Belitsky:2003nz}:
\beq
f_{\gamma^{\; 0}}(\vec{r},x) \propto \text{3-D} \ \text{F.T.} \ \bigg\{ 
{H_q(x,\xi,-\vec{\Delta}^{2}) -\frac{\vec{\Delta}^{2}}{4M_N^2}
E_q(x,\xi,-\vec{\Delta}^{2})} \bigg\},
\eeq
where the expression in brackets could be called the Sachs electric GPD
$G_E(x,\xi,-\vec{\Delta}^{2})$. The function $f_{\gamma^{\; 0}}(\vec{r},x)$ 
is interpreted 
as a 3-D density in the `rest frame' of the proton for quarks with a 
selected value of $x$ and can be defined as a reduction of a so-called 
{quantum phase-space or Wigner distribution}
\cite{Ji:2003ak,Belitsky:2003nz,Belitsky:2003tm}, in the Breit `frame' 
defined as 
\ba
{W_\Gamma(\vec{r},k)} & \equiv & \frac{1}{2M_N} 
\int \frac{d^3 \Delta}{(2\pi)^3}\;
\amp{\vec{\Delta}/2 | \; {\hat {\cal W}_\Gamma(\vec{r},k)} 
\; | -\vec{\Delta}/2}, \\[2 mm]
{\hat {\cal W}_\Gamma(\vec{r},k)} & \equiv & \int d^4 \eta \; 
e^{i\eta \cdot k} \; \overline{\psi}(\vec{r}-\eta/2) \, {\cal L}^\dagger 
\Gamma\, {\cal L} \, \psi(\vec{r} + \eta /2). 
\ea
Fourier transforms of GPDs are obtained as follows (considering 
$\Gamma=\gamma^+$ now):
\ba
f_{\gamma^+}(\vec{r},x) & \equiv & 
{\int \frac{d^2 \bm{k}_\perp}{(2\pi)^2}} 
\left[\int \frac{dk^-}{2\pi} W_{\gamma^+} (\vec{r},k)\right] \\
& & \mbox{} \hspace{-2 cm} \propto \text{F.T.} \bigg\{ 
{H_q(x,\xi,t)} \, 
\bar{u}(\vec{\Delta}/2)\gamma^+
u(-\vec{\Delta}/2)+ {E_q(x,\xi,t)} \, 
\bar{u}(\vec{\Delta}/2) \frac{i \sigma^{+ i} \Delta_i}{2M} 
u(-\vec{\Delta}/2)
\bigg\}. \nn
\ea
Integrating $f_{\gamma^+}(\vec{r},x)$ over the $z$ coordinate yields 
$q(x, \bm{b}_\perp)$ and $\int dx \; f_{\gamma^+}(\vec{r},x)$ is 
the F.T. of the Sachs electric and magnetic form factors.
But {TMPDs can also be seen as reductions of Wigner distributions}\\[-3 mm]
\beq
q(x,\bm{k}_T) = {\int \frac{d^3 r}{(2\pi)^3}}
\left[{\int \frac{dk^-}{2\pi}} W_{\gamma^+} (\vec{r},k)\right]. 
\eeq
Hence, both GPDs and TMPDs can be viewed as different 
reductions of one underlying quantity. Different choices of frame 
lead to complementary physical pictures of these quantities (or their Fourier
transforms), as {momentum and/or spatial distributions}. 

Finally, a suggestion concerning 
the 5-D phase space quantity (which has a straightforward
IMF density interpretation) \\[-3 mm]
\beq
{q(x,\bm{k}_T, \bm{b}_\perp)} \equiv {\int \frac{dz}{2\pi}}
\left[{\int \frac{dk^-}{2\pi}} W_{\gamma^+} (\vec{r},k)\right].
\eeq
Perhaps {$q(x,\bm{k}_T, \bm{b}_\perp)$} 
can be measured in
{hard `semi'-exclusive processes},
such as {$\gamma^* \, p \to V_1 \,
V_2 \, p'$}, with $V_i$ either $\gamma$ or a vector meson and $V_1$ and $V_2$
have a small relative transverse momentum. 
{Note that $\bm{k}_T$ and $\bm{b}_\perp$ are not each other's Fourier
conjugates}. \\[-8 mm]


\begin{theacknowledgments}
I thank the organizers for their kind invitation to present this overview at
the conference and Andrei Belitsky for useful feedback. 
The research of D.B.~has been made possible by 
financial support from the Royal Netherlands Academy of Arts and
Sciences.\\[-8 mm] 

\end{theacknowledgments}


\bibliographystyle{aipproc}   

\bibliography{proc-fb19-hepc}

\IfFileExists{\jobname.bbl}{}
 {\typeout{}
  \typeout{******************************************}
  \typeout{** Please run "bibtex \jobname" to optain}
  \typeout{** the bibliography and then re-run LaTeX}
  \typeout{** twice to fix the references!}
  \typeout{******************************************}
  \typeout{}
 }

\end{document}